\documentclass[twocolumn,tighten]{aastex63}
\hypersetup{linkcolor=red,citecolor=blue,filecolor=cyan,urlcolor=magenta}
\usepackage{amsmath}
\usepackage{natbib}

\newcommand{\degree}{$^{\circ}$}
\newcommand{\loggf}{\mbox{$\log(gf)$}}
\newcommand{\kmsec}{\mbox{km~s$^{\rm -1}$}}
\newcommand{\logg}{\mbox{log~{\it g}}}

\newcommand{\teff}{\mbox{$T_{\rm eff}$}}
\newcommand{\vt}{\mbox{$v_{\rm t}$}}

\newcommand{\logeps}[1]{$\log\varepsilon$(#1)}

\shorttitle{Detection of Al~\textsc{ii} in Metal-Poor Stars}
\shortauthors{Roederer et al.}
\begin{document}

\title{%
Detection of Al~II in the Ultraviolet Spectra of Metal-Poor Stars:\ 
An Empirical LTE Test of NLTE Aluminum Abundance Calculations\footnote{%
Based on observations made with the NASA/ESA 
\textit{Hubble Space Telescope}, 
obtained at the Space Telescope Science Institute, % (STScI),
which is operated by the Association of Universities for 
Research in Astronomy, Inc.\ % (AURA) 
under NASA contract NAS~5-26555.
The data presented in this paper were obtained from the 
Barbara A.\ Mikulski Archive for Space Telescopes. % (MAST).~
%and are associated with programs
%GO-7348,
%GO-7402,
%GO-7433,
%GO-8342,
%GO-9048,
%GO-9455,
%GO-12268,
%GO-12554,
%GO-14161,
%GO-14232,
%GO-14765, and
%GO-15657.
Other data have been obtained from the European Southern Observatory % (ESO) 
Science Archive Facility.
%These data are associated with programs
%66.D-0636(A),
%67.D-0439(A),
%68.B-0475(A),
%68.D-0094(A),
%073.D-0024(A),
%080.D-0347(A), and
%095.D-0504(A).
This research has also made use of the Keck Observatory Archive, % (KOA), 
which is operated by the W.M.\ Keck Observatory and 
the NASA Exoplanet Science Institute, % (NExScI), 
under contract with NASA.~
%These data are associated with programs
%H2aH,
%H39aH, and
%U11H.
This paper includes data taken at The McDonald Observatory
of The University of Texas at Austin,
and the the 6.5~meter 
Magellan Telescopes located at Las Campanas Observatory, Chile.
}
}

\author{Ian U.\ Roederer}
\affiliation{%
Department of Astronomy, University of Michigan,
1085 S.\ University Ave., Ann Arbor, MI 48109, USA}
\affiliation{%
Joint Institute for Nuclear Astrophysics -- Center for the
Evolution of the Elements (JINA-CEE), USA}
\email{Email:\ iur@umich.edu}

\author{James E.\ Lawler}
\affiliation{%
Department of Physics, University of Wisconsin -- Madison,
1150 University Ave., Madison, WI 53706, USA}

\begin{abstract}

We report the detection of an Al~\textsc{ii} line at 2669.155~\AA\
in 11 metal-poor stars, using ultraviolet spectra
obtained with the Space Telescope Imaging Spectrograph
on board the Hubble Space Telescope.
We derive Al abundances from this line using a standard
abundance analysis,
assuming local thermodynamic equilibrium (LTE).~
The mean [Al/Fe] ratio is $-$0.06 $\pm$~0.04 ($\sigma =$ 0.22)
for these 11~stars spanning
$-3.9 <$~[Fe/H]~$< -1.3$,
or [Al/Fe] = $-$0.10 $\pm$~0.04 ($\sigma =$ 0.18)
for 9~stars spanning 
$-3.0 <$~[Fe/H]~$< -1.3$
if two carbon-enhanced stars are excluded.
We use these abundances to perform an empirical test of non-LTE (NLTE)
abundance corrections predicted for resonance lines of Al~\textsc{i},
including the commonly-used optical Al~\textsc{i} line at 3961~\AA.~
The Al~\textsc{ii} line is formed in LTE,
and the abundance derived from this line
matches that derived from high-excitation Al~\textsc{i} lines
predicted to have minimal NLTE corrections.
The differences between the abundance derived from the Al~\textsc{ii} line
and the LTE abundance derived from Al~\textsc{i} resonance lines
are $+$0.4 to $+$0.9~dex, which match 
the predicted NLTE corrections for the Al~\textsc{i} resonance lines.
We conclude that the NLTE abundance calculations 
are approximately correct and
should be applied to LTE abundances derived from
Al~\textsc{i} lines.

\end{abstract}

%%% PLAIN-TEXT ABSTRACT
%We report the detection of an Al II line at 2669.155 Angstroms in 11 metal-poor stars, using ultraviolet spectra obtained with the Space Telescope Imaging Spectrograph on board the Hubble Space Telescope. We derive Al abundances from this line using a standard abundance analysis, assuming local thermodynamic equilibrium (LTE). The mean [Al/Fe] ratio is -0.06 +/- 0.04 (sigma = 0.22) for these 11 stars spanning -3.9 < [Fe/H] < -1.3, or [Al/Fe] = -0.10 +/- 0.04 (sigma = 0.18) for 9 stars spanning -3.0 < [Fe/H] < -1.3 if two carbon-enhanced stars are excluded. We use these abundances to perform an empirical test of non-LTE (NLTE) abundance corrections predicted for resonance lines of Al I, including the commonly-used optical Al I line at 3961 Angstroms. The Al II line is formed in LTE, and the abundance derived from this line matches that derived from high-excitation Al I lines predicted to have minimal NLTE corrections. The differences between the abundance derived from the Al II line and the LTE abundance derived from Al I resonance lines are +0.4 to +0.9 dex, which match the predicted NLTE corrections for the Al I resonance lines. We conclude that the NLTE abundance calculations are approximately correct and should be applied to LTE abundances derived from Al I lines.


\section{Introduction}
\label{intro}

The element aluminum (Al, $Z = 13$) 
is commonly found in stars.
Al is produced mainly during
hydrostatic carbon burning, with additional contributions from
neon burning, in massive stars
(e.g., \citealt{arnett85,thielemann85,woosley95}).
Al is produced during $\alpha$-, $n$-, and $p$-capture reaction sequences 
starting from the neutron-rich seed nuclei $^{22}$Ne and $^{23}$Na.
Theoretically Al has some characteristics of a ``primary'' element,
in that the [Al/Fe] ratios are independent of the
initial metallicities of the stars where Al is produced,
because its seed nuclei are themselves produced during
the normal course of stellar burning in massive stars.
Al also has a ``secondary'' metallicity dependence, 
because these seed nuclei may also be present initially,
thus amplifying Al production.
In Galactic chemical evolution models,
the [Al/Fe] ratio is generally flat or shows a mild increase
with increasing [Fe/H]
(e.g., \citealt{timmes95,kobayashi06,kobayashi20,romano10}).

Unfortunately,
the method of analyzing Al lines in stellar spectra 
greatly influences the derived abundances and
thus our understanding of Al nucleosynthesis.
Al abundances in metal-poor stars were initially found to exhibit a 
strong secondary behavior,
sub-solar at low metallicity and approaching or even exceededing the
solar ratio at higher metallicity
(e.g., \citealt{peterson78,magain87,gratton88,
mcwilliam95b,ryan96,chen00}).
Those results were based on the assumption that
the excitation and ionization
could be adequately described by
local thermodynamic equilibrium (LTE).~
Applying non-LTE (NLTE) calculations,
in contrast, yields
[Al/Fe] ratios much closer to the solar ratio across a wide range
of metallicities ($-4 <$ [Fe/H] $< 0$;
e.g., \citealt{baumueller97,gehren04,andrievsky08}).
Consequently, the NLTE Al abundances
better match the theoretically-predicted behavior.

The choice of which Al spectral lines to use 
also has an outsized influence on the derived Al abundances.
It has long been discussed 
(e.g., \citealt{francois86,gratton88,ryan96})
whether high-excitation
Al~\textsc{i} lines in the red 
(e.g., $\lambda$6696, 6698~\AA)
yield abundances higher than
the resonance Al~\textsc{i} lines in the blue
($\lambda$3944, 3961~\AA).~
Observational errors were once large enough to 
sustain this debate.
Modern analyses 
(e.g., \citealt{mashonkina16al,zhao16,nordlander17al})
have shown conclusively that
the high-excitation lines yield abundances higher by
several tenths of a dex than the resonance lines.
Several sets of high-excitation Al~\textsc{i} lines yield
reliable abundances in LTE, yet
only the blue resonance lines are detected
in warm or metal-poor stars.
This situation presents a challenge for interpreting
Al abundances in stars spanning a wide range of temperatures
and metallicities.

We have identified an Al~\textsc{ii} 
absorption line at 2669.155~\AA\ in the ultraviolet (UV) spectra of 
several metal-poor stars.
This line is relatively unblended,
and to the best of our knowledge it has not been used previously
to derive Al abundances.
It arises from the 2p$^{6}$3s ground state of Al$^{+}$,
and it is the only line connected to the Al$^{+}$ ground level
that is detectable in near-UV, optical,
or near-infrared spectra.

The first ionization potential of Al is low, 5.99~eV,
and the second ionization potential is high, 18.83~eV,
so Al$^{+}$
is the dominant ionization state of Al in the atmospheres
of metal-poor stars.
Abundances derived from the UV Al~\textsc{ii} line,
which should be formed in LTE
\citep{mashonkina16al}, can be used 
to empirically assess the deviations from LTE
of transitions in neutral Al.
Similar tests have been performed previously for other elements.
The most extensive studies have focused on iron-group elements
(e.g., \citealt{sneden16,roederer18a,cowan20}).
Some have examined
the $\alpha$ elements Mg \citep{spite17}
and Ca \citep{denhartog21b},
while others have examined neutron-capture elements
(e.g., \citealt{roederer12b,peterson20,roederer20}).
UV spectra collected with the
Hubble Space Telescope (HST) are crucial to this work because
many of the lines arising in ionized atoms are
found in the UV, below the atmospheric cutoff near 3000~\AA.~

In this paper, we present an analysis of Al~\textsc{i} and \textsc{ii} lines
in metal-poor stars.
We compare Al abundances derived from 
UV and optical Al~\textsc{i} and \textsc{ii} lines
as a test of the NLTE predictions, and we
present [Al/Fe] ratios for this sample of metal-poor stars.
We define the Al abundance as
\logeps{Al}~$\equiv \log_{10}(N_{\rm Al}/N_{\rm H})+$12.0.
We define the abundance ratio of Al and Fe relative to the
Solar ratio as
[Al/Fe] $\equiv \log_{10} (N_{\rm Al}/N_{\rm Fe}) - \log_{10} (N_{\rm Al}/N_{\rm Fe})_{\odot}$, where
\logeps{Al}$_{\odot} =$ 6.45 and \logeps{Fe}$_{\odot} =$ 7.50
\citep{asplund09}.
By convention, abundances or ratios denoted 
with the ionization state are understood to be 
the total elemental abundance as derived from transitions of that
particular ionization state 
after Saha ionization corrections have been applied.

\section{Atomic Data}
\label{alatomic}

The one stable isotope of Al, $^{27}$Al, has non-zero nuclear spin,
$I = 5/2$, which produces 
hyperfine structure (HFS) splitting
in the electronic energy levels.
We reconstruct the HFS splitting
of the Al~\textsc{ii} line at 2669~\AA\ using the HFS
magnetic dipole, $A$, 
and 
electric quadrupole, $B$, constants 
for the upper 3s3p $^{3}$P$^{\rm o}_{1}$ level;
these values are 0 for the lower 2p$^{6}$3s$^{2}$ $^{1}$S$_{0}$ level.
We calculate the HFS $A$ and $B$ values that reproduce the
splittings measured by \citet{itano07},
$A = 1339.311$~MHz and
$B = -21.606$~MHz.
Two measured splittings and two adjustable constants yield a perfect fit. 
However, \citeauthor{itano07} 
mention that the second-order magnetic dipole energy 
is comparable to the first-order electric quadrupole energy.
Any comparison of these HFS constants to ab-initio theory 
requires some care, especially for the
HFS $B$ constant. 
The above $A$ value agrees to better than 1\% 
with the ab-initio theoretical result from \citet{zhang17al},
$A = 1327.3 \pm 10.2$~MHz. 
The above $B$ value does not agree as well with the result from 
\citeauthor{zhang17al},
$B = -15.1 \pm 0.2$~MHz, 
but $B$ values are generally not important in astrophysical research. 
There is no doubt that the splitting measurements from 
\citeauthor{itano07}, made to kHZ accuracy, are better than
theoretical values for the HFS $A$ and $B$.
We normalize the HFS line component shifts to the center-of-gravity 
wavelength calculated from the 
National Institute of Standards and Technology (NIST) energy levels.
The relative strengths of each component are calculated 
using the $LS$ angular momentum coupling formulae 
presented in \citet{condon35}.
Table~\ref{alhfstab} presents the HFS line component pattern.

\begin{deluxetable*}{ccccccc}
\tablecaption{Hyperfine Structure for the 
$^{27}$Al~\textsc{ii} Line at 2669~\AA\
\label{alhfstab}}
%\tablewidth{0pt}
\tabletypesize{\small}
\tablehead{
\colhead{Wavenumber} &
\colhead{$\lambda_{\rm air}$} &
\colhead{$F_{\rm upper}$} &
\colhead{$F_{\rm lower}$} &
\colhead{Component Position} &
\colhead{Component Position} &
\colhead{} \\
\colhead{(cm$^{-1}$)} &
\colhead{(\AA)} &
\colhead{} &
\colhead{} &
\colhead{(cm$^{-1}$)} &
\colhead{(\AA)} &
\colhead{Strength} 
}
\startdata
37453.91 & 2669.155 & 3.5 & 2.5 & $+$0.1115 & $-$0.00795 & 0.4444 \\
37453.91 & 2669.155 & 2.5 & 2.5 & $-$0.0441 & $+$0.00314 & 0.3333 \\
37453.91 & 2669.155 & 1.5 & 2.5 & $-$0.1569 & $+$0.01118 & 0.2222 \\
\enddata
\tablecomments{%
Energy levels from the NIST ASD and the index of air \citep{peck72}
are used to compute the
center-of-gravity wavenumbers and air wavelengths, $\lambda_{\rm air}$.
Line component positions are given relative to those values.
%Table~\ref{alhfstab} is available in the online edition
%of the journal in machine-readable format.
}
\end{deluxetable*}

The NIST Atomic Spectra Database (ASD, version 5.8; \citealt{kramida20}) 
recommends a \loggf\ value of $-$4.979 for the
Al~\textsc{ii} line at 2669~\AA\ \citep{trabert99},
with a grade of ``A+'' (uncertainty $<$~2\%).
We adopt this value.

We also examine 15 Al~\textsc{i} lines.
The UV Al~\textsc{i} lines were selected for analysis based on
their detectability and relative lack of blends in the spectra of
\object[HD 84937]{HD~84937} and 
\object[HD 222925]{HD~222925}
(see Section~\ref{sample}).
For each line, Table~\ref{atomictab} lists the wavelength,
excitation potential (E.P.) of the lower level,
\loggf\ value, NIST ASD grade on the accuracy of the \loggf\ value,
and source of the HFS line component pattern.
HFS patterns are available for nine of these lines in the
Vienna Atomic Line Database
(VALD3; \citealt{piskunov95,pakhomov19}),
which makes use of data presented in
\citet{stuck70al},
\citet{falkenberg79al},
\citet{zhankui82al},
\citet{belfrage84al},
\citet{jonsson84al},
\citet{biemont87al},
\citet{chang90al},
and
\citet{brown99al}.
We adopt the NIST ASD recommended \loggf\ values for the Al~\textsc{i} lines,
which are based on data presented in
\citet{davidson90},
\citet{mendoza95}, 
and
\citet{hannaford99}.
We include damping constants for
collisional broadening with neutral hydrogen
from \citet{barklem00neutral}, when available,
otherwise we adopt the broadening as 
described by the \citet{unsold55} recipe
in our spectrum synthesis calculations.

\begin{deluxetable}{cccccc}
\tablecaption{Atomic Data for Lines Used in This Study
\label{atomictab}}
%\tablewidth{0pt}
\tabletypesize{\small}
\tablehead{
\colhead{Species} &
\colhead{$\lambda$} &
\colhead{E.P.} &
\colhead{\loggf\tablenotemark{a}} &
\colhead{Grade\tablenotemark{b}} &
\colhead{HFS\tablenotemark{a}} \\
\colhead{} &
\colhead{(\AA)} &
\colhead{(eV)} &
\colhead{} &
\colhead{} &
\colhead{} 
}
\startdata
\mbox{Al~\textsc{i}}  & 2118.332 & 0.00 & $-$1.56 & C  & \nodata \\
\mbox{Al~\textsc{i}}  & 2129.678 & 0.00 & $-$1.38 & C  & \nodata \\
\mbox{Al~\textsc{i}}  & 2199.180 & 0.00 & $-$2.60 & C  & VALD \\
\mbox{Al~\textsc{i}}  & 2204.660\tablenotemark{c} 
                                 & 0.00 & $-$0.90 & C+ & \nodata \\
\mbox{Al~\textsc{i}}  & 2263.738 & 0.01 & $-$1.94 & C  & VALD \\
\mbox{Al~\textsc{i}}  & 2269.220 & 0.01 & $-$1.41 & C+ & \nodata \\
\mbox{Al~\textsc{i}}  & 2372.070 & 0.00 & $-$2.01 & C  & VALD \\
\mbox{Al~\textsc{i}}  & 2373.122 & 0.01 & $-$0.34 & B  & VALD \\
\mbox{Al~\textsc{i}}  & 2567.982 & 0.00 & $-$1.12 & B  & VALD \\
\mbox{Al~\textsc{i}}  & 3944.006 & 0.00 & $-$0.64 & B+ & VALD \\
\mbox{Al~\textsc{i}}  & 3961.519 & 0.01 & $-$0.33 & B+ & VALD \\
\mbox{Al~\textsc{i}}  & 6696.019 & 3.14 & $-$1.57 & C+ & VALD \\
\mbox{Al~\textsc{i}}  & 6698.670 & 3.14 & $-$1.87 & C+ & VALD \\
\mbox{Al~\textsc{i}}  & 7835.309 & 4.02 & $-$0.69 & B  & \nodata \\
\mbox{Al~\textsc{i}}  & 7836.134\tablenotemark{d}
                                 & 4.02 & $-$0.53 & B+ & \nodata \\
\hline
\mbox{Al~\textsc{ii}} & 2669.155 & 0.00 & $-$4.98 & A+ & This study \\
\enddata
\tablenotetext{a}{See text for references.}
\tablenotetext{b}{\loggf\ grade assigned in the NIST ASD
(A+ = 2\%,
A =   3\%,
B+ =  7\%,
B =  10\%,
C+ = 18\%,
C =  25\%).
}
\tablenotetext{c}{Line blends together with the weaker
Al~\textsc{i} line at 2204.619~\AA, with 
E.P.\ = 0.01~eV, \loggf\ = $-$2.29, grade ``C,''
and HFS from VALD.}
\tablenotetext{d}{Line blends together with the weaker
Al~\textsc{i} line at 7836.134~\AA, with
E.P.\ = 4.02~eV, \loggf\ = $-$1.83, and grade ``C+.''}
%\tablecomments{%}
\end{deluxetable}

\section{Archival Observations}
\label{sample}

We collect archival UV and optical spectra for 11 stars
where the Al~\textsc{ii} line is detected and useful for analysis.
Table~\ref{obstab} lists the basic characteristics
of these spectra, including the instrument used,
program identification (ID) number, 
datasets,
original principle investigator (PI) of the observations,
wavelength ($\lambda$) coverage, and
spectral resolving power ($R \equiv \lambda/\delta\lambda$).
The Space Telescope Imaging Spectrograph 
(STIS; \citealt{kimble98,woodgate98}) spectra
were obtained through the 
Mikulski Archive for Space Telescopes (MAST)
and processed automatically by the CALSTIS pipeline.
The High Accuracy Radial velocity Planet Searcher 
(HARPS; \citealt{mayor03}) spectra and 
Ultraviolet Visual Echelle Spectrograph 
(UVES; \citealt{dekker00}) spectra
were obtained through the European Southern Observatory % (ESO)
Science Archive Facility.
The High Resolution Echelle Spectrometer
(HIRES; \citealt{vogt94}) spectra
were obtained through the Keck Observatory Archives. % (KOA).
The Magellan Inamori Kyocera Echelle
(MIKE; \citealt{bernstein03}) spectra
and
Robert G.\ Tull Coud\'{e} \citep{tull95} spectra
have been collected over the years by us.
We use the Image Reduction and Analysis Facility (IRAF; \citealt{tody93})
software to shift to rest velocity, co-add, and
continuum normalize the spectra.

\begin{figure*}
\begin{center}
\includegraphics[angle=0,width=6.5in]{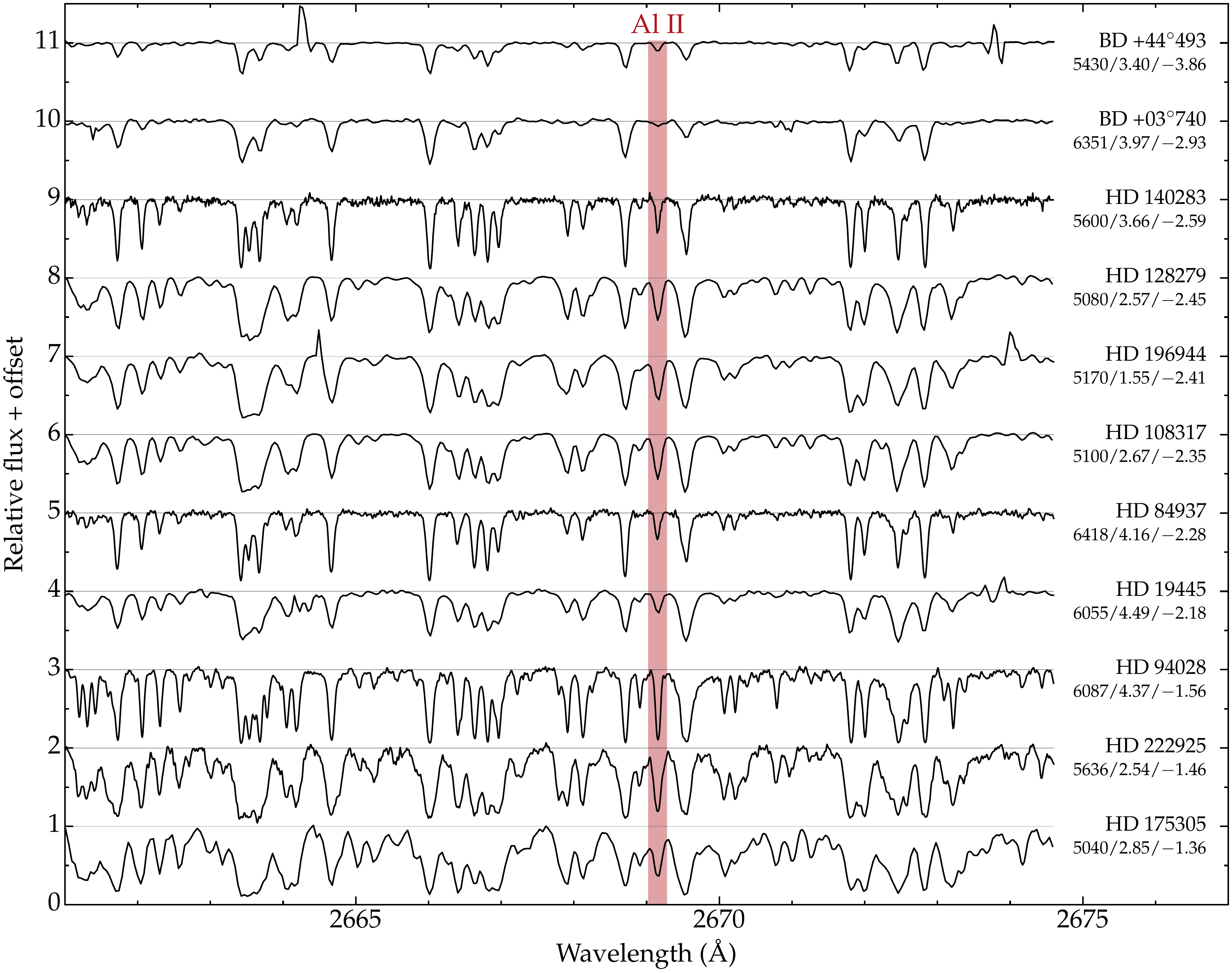}
\end{center}
\caption{
\label{plot2669}
Sections of the STIS spectra.
The Al~\textsc{ii} line at 2669.155~\AA\ is marked by the shaded box.
Thin gray lines mark the approximate continuum in each star.
The spectra have been shifted vertically for display purposes,
and they are ordered by increasing [Fe/H] from top to bottom.
The star names and \teff/\logg/[Fe/H] values
(in units of K, $\log$~cm~s$^{-2}$, and dex)
are listed.
}
\end{figure*}

Figure~\ref{plot2669} illustrates a region around the 
Al~\textsc{ii} line in the STIS spectra
of all 11~stars in our sample.
Typical signal-to-noise (S/N) ratios
near the Al~\textsc{ii} line 
range from 40 to 70~pix$^{-1}$, 
except for the spectra of 
\object[HD 108317]{HD 108317} and
\object[HD 128279]{HD 128279},
where S/N $\approx$~160~pix$^{-1}$.
Figure~\ref{plot2669} illustrates that 
this line remains largely free of blends, even in the coolest 
and most metal-rich stars in the sample.
The Al~\textsc{ii} line with the lowest detection significance,
3.5$\sigma$, is found in 
\object[BD+03 740]{BD~$+$03$^{\circ}$740}.
All other detections are 8$\sigma$ significance or greater.

S/N ratios around Al~\textsc{i} lines are
$\gtrsim$~30~pix$^{-1}$ at 2200~\AA\ and 
increase to longer wavelengths.
More extensive lists of
S/N ratios at a variety of UV and optical wavelengths
for these spectra can be found in 
Table~1 of \citet{roederer12c},
Table~2 of \citet{roederer12d},
Section~2 of \citet{roederer14d},
Table~1 of \citet{roederer18b},
and
Section~3 of \citet{peterson20}.

\startlongtable
\begin{deluxetable*}{lcccccr}
\tablecaption{Log of Observations
\label{obstab}}
%\tablewidth{0pt}
\tabletypesize{\small}
%\tablewidth{6.5in}
\tablehead{
\colhead{Star} &
\colhead{Instrument} &
\colhead{Program ID} &
\colhead{MAST Datasets} &
\colhead{PI} &
\colhead{$\lambda$ (\AA)} &
\colhead{$R$}
}
\startdata
\mbox{BD $+$03\degree740}  & STIS  & GO-14232 & OCTS01010-2030      & Roederer     & 2278--3068 &  30,000 \\
                           & MIKE  & \tablenotemark{a} & \nodata    & Roederer     & 3340--9410 &  42,000 \\
\hline
\mbox{BD $+$44\degree493}  & STIS  & GO-12554 & OBQ603010-4040      & Beers        & 2278--3073 &  30,000 \\
                           & Tull  & \tablenotemark{b} & \nodata    & Roederer     & 3636--8000 &  33,000 \\
\hline
\mbox{HD 19445}            & STIS  & GO-7402  & O56D01010-3010      & Peterson     & 2313--3067 &  30,000 \\
                           & UVES  & 66.D-0636(A) &      \nodata    & Piotto       & 3760--4980 &  41,000 \\
                           & UVES  & 68.D-0094(A) &      \nodata    & Primas       & 5841--6810 &  51,000 \\
\hline
\mbox{HD 84937}            & STIS  & GO-14161 & OCTKA0010-D01030    & Peterson     & 1879--3143 & 114,000 \\
%                          & STIS  & GO-7402  & O56D04010-5010      & Peterson     & 2279--3117 &  30,000 \\
                           & UVES  & 073.D-0024(A) &     \nodata    & Akerman      & 3757--4980 &  54,000 \\
                           & UVES  & 266.D-5655(A) &     \nodata &\tablenotemark{c}& 5836--6809 &  74,000 \\
\hline
\mbox{HD 94028}            & STIS  & GO-8179  & O5CN01010-3040      & Duncan       & 1879--2148 & 114,000 \\
                           & STIS  & GO-14161 & OCTKB0010-6030      & Peterson     & 2128--3143 & 114,000 \\
%                          & STIS  & GO-7402  & O56D06010           & Peterson     & 2280--3117 &  30,000 \\
                           & Tull  & \tablenotemark{b} & \nodata    & Roederer     & 3647--8000 &  33,000 \\
\hline
\mbox{HD 108317}           & STIS  & GO-12268 & OBJQ01010-3050      & Roederer     & 2280--3115 &  30,000 \\
                           & STIS  & GO-12976 & OBXV01010-4030      & Roederer     & 1610--2365 &  30,000 \\
                           & MIKE  & \tablenotemark{b} & \nodata    & Thompson     & 3340--8000 &  41,000 \\
\hline
\mbox{HD 128279}           & STIS  & GO-12268 & OBJQ04010-6050      & Roederer     & 2280--3115 &  30,000 \\
                           & STIS  & GO-12976 & OBXV05010-7050      & Roederer     & 1610--2365 &  30,000 \\
                           & MIKE  & \tablenotemark{b} & \nodata    & Thompson     & 3340--8000 &  41,000 \\
\hline
\mbox{HD 140283}           & STIS  & GO-7348  & O55Z01010-2070      & Edvardsson   & 1932--2212 & 114,000 \\
                           & STIS  & GO-9455  & O6LM71010-40        & Peterson     & 2390--3140 &  51,000 \\
                           & HARPS & 080.D-0347(A) &     \nodata    & Heiter       & 3784--6917 & 115,000 \\
\hline
\mbox{HD 175305}           & STIS  & GO-8342  & O5F609010\tablenotemark{d} & Cowan & 2277--3119 &  30,000 \\
                           & Tull  & \tablenotemark{b} & \nodata    & Roederer     & 3679--8000 &  33,000 \\
\hline
\mbox{HD 196944}           & STIS  & GO-12554 & OBQ601010-30        & Beers        & 2278--3073 &  30,000 \\
                           & STIS  & GO-14765 & OD5A01010-14010     & Roederer     & 2029--2303 & 114,000 \\
                           & MIKE  & \tablenotemark{b} & \nodata    & Roederer     & 3350--9150 &  41,000 \\
\hline
\mbox{HD 222925}           & STIS  & GO-15657 & ODX901010-60030     & Roederer     & 1936--3145 & 114,000 \\
                           & MIKE  & \tablenotemark{e} & \nodata    & Roederer     & 3330--9410 &  66,000 \\
\enddata
\tablenotetext{a}{Previously unpublished spectrum collected 2019 October 24}
\tablenotetext{b}{See \citet{roederer14c}}
\tablenotetext{c}{UVES Paranal Observatory Project \citep{bagnulo03}}
\tablenotetext{d}{Via StarCat \citep{ayres10}}
\tablenotetext{e}{See \citet{roederer18c}}
\end{deluxetable*}

\section{Analysis}
\label{analysis}

\subsection{Model Atmospheres}
\label{model}

We adopt the model parameters
(effective temperature, \teff;
log of the surface gravity, \logg;
microturbulent velocity parameter, \vt;
and
model metallicity, [M/H])
derived previously for these stars.
We interpolate model atmospheres from the 1D,
$\alpha$-enhanced
ATLAS9 grid of models \citep{castelli04}.
These values are listed in Table~\ref{atmtab}.

\begin{deluxetable*}{lcccccc}
\tablecaption{Model Atmosphere Parameters
\label{atmtab}}
%\tablewidth{0pt}
\tabletypesize{\small}
\tablehead{
\colhead{Star} &
\colhead{\teff} &
\colhead{\logg} &
\colhead{\vt} &
\colhead{[M/H]} &
\colhead{[Fe/H]\tablenotemark{a}} &
\colhead{References} \\
\colhead{} &
\colhead{(K)} &
\colhead{[cgs]} &
\colhead{(\kmsec)} &
\colhead{} &
\colhead{} &
\colhead{}
}
\startdata
BD $+$03\degree740  & 6351 & 3.97 & 1.70 & $-$2.90 & $-$2.93 & \citet{roederer18b} \\
BD $+$44\degree493  & 5430 & 3.40 & 1.30 & $-$3.80 & $-$3.86 & \citet{ito13,placco14bdp44} \\
%BD $-$13\degree3442 & 6405 & 4.04 & 1.60 & $-$2.85 & $-$2.85 & \citet{roederer18b} \\
%CD $-$33\degree1173 & 6625 & 4.29 & 1.60 & $-$3.00 & $-$3.10 & \citet{roederer18b} \\
%HD 2454             & 6050 & 4.04 & 1.20 & $-$0.44 & $-$0.35 & \citet{roederer12c} \\
HD 19445            & 6055 & 4.49 & 1.20 & $-$2.20 & $-$2.18 & \citet{roederer18b} \\
HD 84937            & 6418 & 4.16 & 1.50 & $-$2.25 & $-$2.28 & \citet{roederer18b} \\
HD 94028            & 6087 & 4.37 & 1.10 & $-$1.60 & $-$1.56 & \citet{roederer18b} \\
%HD 107113           & 6060 & 3.99 & 1.25 & $-$0.48 & $-$0.43 & \citet{roederer12c} \\
HD 108317           & 5100 & 2.67 & 1.50 & $-$2.37 & $-$2.35 & \citet{roederer12d} \\
HD 128279           & 5080 & 2.57 & 1.60 & $-$2.46 & $-$2.45 & \citet{roederer12d} \\
HD 140283           & 5600 & 3.66 & 1.15 & $-$2.62 & $-$2.59 & \citet{roederer12c} \\
HD 175305           & 5040 & 2.85 & 2.00 & $-$1.48 & $-$1.36 & \citet{cowan05} \\
HD 196944           & 5170 & 1.60 & 1.55 & $-$2.41 & $-$2.41 & \citet{placco15cemps} \\
HD 222925           & 5636 & 2.54 & 2.20 & $-$1.50 & $-$1.46 & \citet{roederer18c} \\
\enddata
\tablenotetext{a}{[Fe/H] derived from Fe~\textsc{ii} lines on the \loggf\ scale
described in the text.}
%\tablecomments{%}
\end{deluxetable*}

\subsection{Fe Abundances}
\label{iron}

We derive abundances using a recent version of the MOOG
line analysis software (\citealt{sneden73}, 2017 version).
MOOG assumes that LTE
holds in the line-forming layers of the atmosphere.
This version of MOOG accounts for Rayleigh scattering,
which affects the continuous opacity at shorter wavelengths,
as isotropic, coherent scattering \citep{sobeck11}.

We derive Fe abundances using Fe~\textsc{ii} lines,
which are less susceptible than Fe~\textsc{i} lines to departures from LTE.~
We update previously-derived abundances to the
Fe~\textsc{ii} \loggf\ scale established
by \citet{denhartog19}.
If an Fe~\textsc{ii} line is not present in the \citeauthor{denhartog19}\
list, we adopt the \loggf\ value from \citet{melendez09fe},
if available, otherwise we default to the recommended NIST value.
We discard any Fe~\textsc{ii} lines
in the ``Balmer Dip'' region 
($\approx$3100--3700~\AA), 
where multiple 1D LTE line analysis codes yield abundances
that are systematically different from those derived using
lines at shorter and longer wavelengths \citep{roederer18b}.
Our updated [Fe/H] ratios are listed in Table~\ref{atmtab}.
The mean changes are $\leq +0.05$~dex for all stars except
\object[HD 94028]{HD~94028}, for which the change is $+0.09$~dex.
These small changes would have had a negligible impact on the 
model atmospheres derived previously.
Nearly all previous studies derived \logg, the one
parameter potentially most sensitive to Fe~\textsc{ii} lines,
using distances from parallax measurements
rather than Fe ionization balance.
The two stars whose \logg\ values were derived using 
the Fe ionization balance method,
\object[BD+44 493]{BD~$+$44$^{\circ}$493}
and 
\object[HD 196944]{HD~196944},
are unaffected because their updated Fe abundances changed by
$\leq 0.01$~dex.

\subsection{Al Abundances}
\label{aluminum}

\begin{figure*}
\begin{center}
\includegraphics[angle=0,width=6.5in]{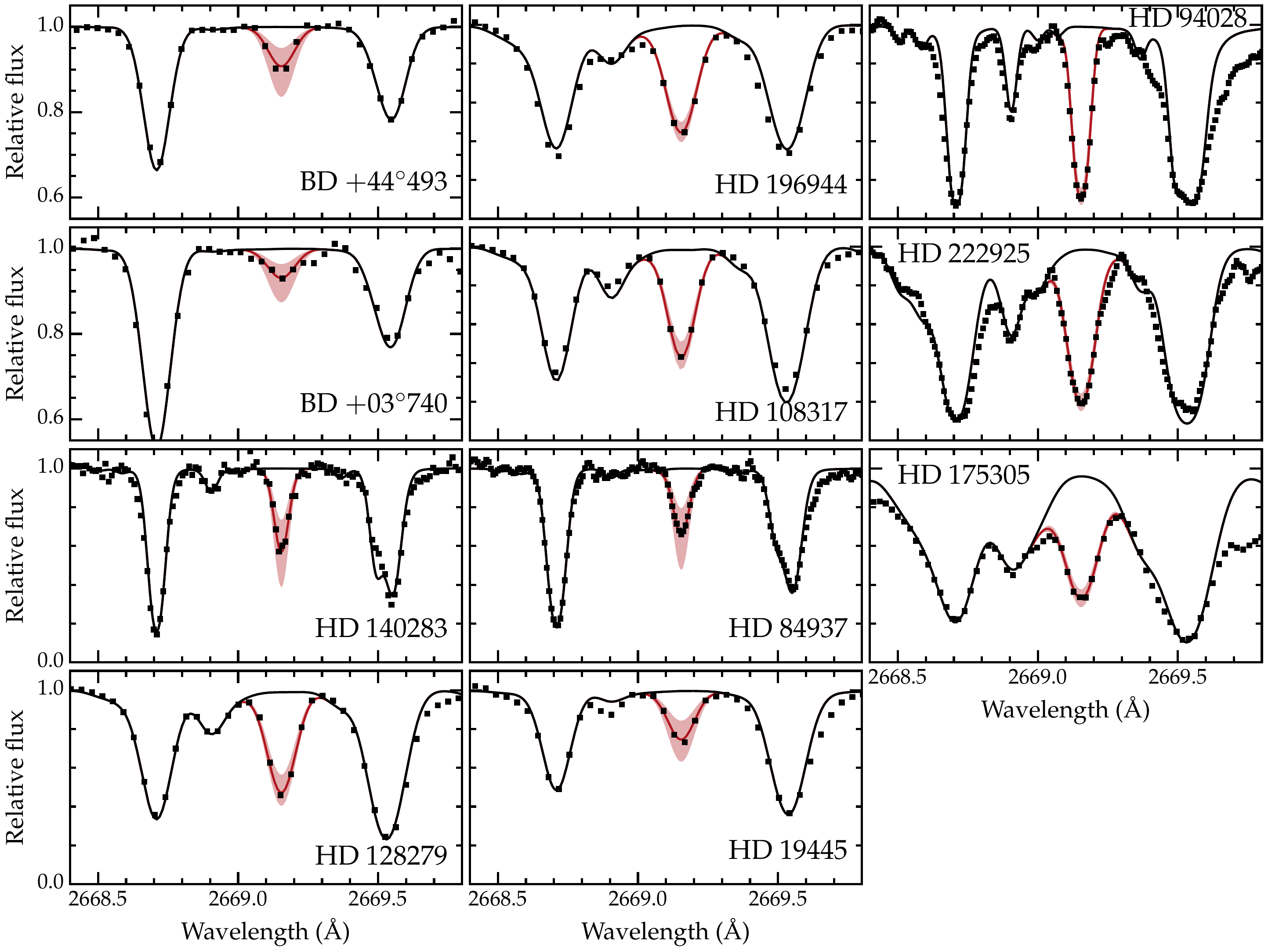}
\end{center}
\caption{
\label{synthplot2669}
Comparison of synthetic and observed spectra around the Al~\textsc{ii} line.
The filled dots mark the observed spectra.
The red solid lines represent synthetic spectra with the 
best-fit Al abundance,
and the pink bands represent changes in this abundance
by factors of $\pm$~2 (0.3~dex).
The black lines represent synthetic spectra with no Al.
The panels are ordered by increasing [Fe/H].
}
\end{figure*}

We derive Al abundances by matching synthetic spectra to the
observed spectra.
We generate line lists for synthesis with our own version
of the `linemake' software\footnote{%
\url{https://github.com/vmplacco/linemake}},
which includes updates relevant to the UV spectral range.
Figure~\ref{synthplot2669} illustrates these fits
to the Al~\textsc{ii} line.
This line remains on the linear part of the curve-of-growth
in most stars in our sample.
We also derive Al abundances from Al~\textsc{i} lines.
Not all Al~\textsc{i} lines are detectable or useful
as abundance indicators in each star.
The lines from the ground level, while nearly always present
in these spectra, are frequently saturated or blended
with other species.
The Al~\textsc{i} line at 3944~\AA, for example,
is frequently blended with
CH lines \citep{arpigny83}, 
and it is useful as an abundance indicator
only in a few warm, metal-poor stars without enhanced carbon.
The high-excitation lines are only detectable in
the most metal-rich stars in our sample.
Table~\ref{lineabundtab} lists the abundance derived
from each line.

Following
\citet{roederer18c},
we compute 1$\sigma$ abundance uncertainties as follows.
We draw 250 resamples of the stellar parameters
(with assumed 
uncertainties of 100~K in \teff,
0.2~dex in \logg,
0.2~\kmsec\ in \vt, and
0.2~dex in [M/H]),
\loggf\ values
(with uncertainties quoted in the original source references or NIST ASD),
and equivalent widths approximated from 
the abundance derived via synthesis using a reverse
curve-of-growth method
(with assumed 5\%, or minimum 2~m\AA, uncertainties).
We recompute the $\log\varepsilon$ Al and Fe abundances and the
[Al/Fe] ratio for each resample.
Each distribution is roughly symmetric, and we adopt the 16th 
and 84th percentiles as the 1$\sigma$ uncertainty on each quantity.

\begin{deluxetable*}{lcccccccccccccccc}
\tablecaption{Abundances Derived from Individual Lines
\label{lineabundtab}}
%\tablewidth{0pt}
\tabletypesize{\small}
\tablehead{
\colhead{Star} &
\colhead{Al~\textsc{i}} &
\colhead{Al~\textsc{i}} &
\colhead{Al~\textsc{i}} &
\colhead{Al~\textsc{i}} &
\colhead{Al~\textsc{i}} &
\colhead{Al~\textsc{i}} &
\colhead{Al~\textsc{i}} &
\colhead{Al~\textsc{i}} &
\colhead{Al~\textsc{i}} &
\colhead{Al~\textsc{i}} &
\colhead{Al~\textsc{i}} &
\colhead{Al~\textsc{i}} &
\colhead{Al~\textsc{i}} &
\colhead{Al~\textsc{i}} &
\colhead{Al~\textsc{i}} &
\colhead{Al~\textsc{ii}} \\
\colhead{} &
\colhead{2118} &
\colhead{2129} &
\colhead{2199} &
\colhead{2204} &
\colhead{2263} &
\colhead{2269} &
\colhead{2372} &
\colhead{2373} &
\colhead{2567} &
\colhead{3944} &
\colhead{3961} &
\colhead{6696} &
\colhead{6698} &
\colhead{7835} &
\colhead{7836} &
\colhead{2669} }
\startdata
\hline
\multicolumn{17}{c}{$\log\varepsilon$ abundance} \\
% & $\log\varepsilon$ & $\log\varepsilon$ & $\log\varepsilon$ & $\log\varepsilon$ & $\log\varepsilon$ & $\log\varepsilon$ & $\log\varepsilon$ &
%   $\log\varepsilon$ & $\log\varepsilon$ & $\log\varepsilon$ & $\log\varepsilon$ & $\log\varepsilon$ & $\log\varepsilon$ & $\log\varepsilon$ &
%   $\log\varepsilon$ & $\log\varepsilon$ \\
\hline
BD $+$03\degree740  &\nodata&\nodata&\nodata&\nodata&\nodata&\nodata&\nodata& 2.90  & 3.02  & 3.00  & 2.73  &\nodata&\nodata&\nodata&\nodata& 3.52 \\
BD $+$44\degree493  &\nodata&\nodata&\nodata&\nodata&\nodata&\nodata&\nodata& 2.02  & 2.22  &\nodata& 2.05  &\nodata&\nodata&\nodata&\nodata& 2.92 \\
HD 19445            &\nodata&\nodata&\nodata&\nodata&\nodata&\nodata& 3.94  & 3.87  & 3.50  & 3.72  &\nodata&\nodata&\nodata&\nodata&\nodata& 4.25 \\
HD 84937            & 3.53  & 3.65  &\nodata& 3.65  & 3.63  & 3.68  & 3.80  & 3.42  & 3.59  & 3.50  &\nodata&\nodata&\nodata&\nodata&\nodata& 4.20 \\
HD 94028            & 4.83  &\nodata& 4.68  &\nodata& 4.60  & 4.74  & 4.84  &\nodata& 4.68  &\nodata&\nodata&\nodata&\nodata&\nodata&\nodata& 5.18 \\
HD 108317           & 3.33  &\nodata& 3.19  & 3.20  & 3.25  & 3.33  & 3.22  &\nodata&\nodata&\nodata& 3.22  &\nodata&\nodata&\nodata&\nodata& 3.90 \\
HD 128279           &\nodata&\nodata& 3.15  & 3.36  & 3.33  & 3.40  & 3.21  &\nodata&\nodata&\nodata& 3.16  &\nodata&\nodata&\nodata&\nodata& 3.72 \\
HD 140283           & 2.87  & 2.94  & 3.27  & 3.02  & 2.69  & 3.00  &\nodata&\nodata& 3.00  &\nodata& 2.83  &\nodata&\nodata&\nodata&\nodata& 3.65 \\
HD 175305           &\nodata&\nodata&\nodata&\nodata&\nodata&\nodata&\nodata&\nodata&\nodata&\nodata& 4.21  &\nodata&\nodata& 5.01  & 4.95  & 4.76 \\
HD 196944           & 3.20  & 3.27  &\nodata& 3.21  & 3.16  & 3.22  & 3.28  &\nodata&\nodata&\nodata&\nodata&\nodata&\nodata&\nodata&\nodata& 4.16 \\
HD 222925           & 4.32  & 4.28  & 4.27  & 4.21  & 4.12  & 4.27  & 4.17  &\nodata& 4.20  &\nodata&\nodata& 4.87  & 4.90  &\nodata& 4.63  & 4.77 \\
\hline\hline
\multicolumn{17}{c}{Fitting uncertainty in $\log\varepsilon$ abundance} \\
% & Unc. & Unc. & Unc. & Unc. & Unc. & Unc. & Unc. & Unc. & Unc. & Unc. & Unc. & Unc. & Unc. & Unc. & Unc. & Unc. \\
\hline
BD $+$03\degree740  &\nodata&\nodata&\nodata&\nodata&\nodata&\nodata&\nodata& 0.10  & 0.15  & 0.10  & 0.15  &\nodata&\nodata&\nodata&\nodata& 0.10 \\
BD $+$44\degree493  &\nodata&\nodata&\nodata&\nodata&\nodata&\nodata&\nodata& 0.10  & 0.10  &\nodata& 0.10  &\nodata&\nodata&\nodata&\nodata& 0.05 \\
HD 19445            &\nodata&\nodata&\nodata&\nodata&\nodata&\nodata& 0.15  & 0.15  & 0.15  & 0.15  &\nodata&\nodata&\nodata&\nodata&\nodata& 0.05 \\
HD 84937            & 0.10  & 0.10  &\nodata& 0.10  & 0.15  & 0.10  & 0.15  & 0.15  & 0.10  & 0.05  &\nodata&\nodata&\nodata&\nodata&\nodata& 0.05 \\
HD 94028            & 0.20  &\nodata& 0.15  &\nodata& 0.10  & 0.15  & 0.10  &\nodata& 0.15  &\nodata&\nodata&\nodata&\nodata&\nodata&\nodata& 0.20 \\
HD 108317           & 0.15  &\nodata& 0.25  & 0.15  & 0.15  & 0.20  & 0.15  &\nodata&\nodata&\nodata& 0.15  &\nodata&\nodata&\nodata&\nodata& 0.15 \\
HD 128279           &\nodata&\nodata& 0.25  & 0.15  & 0.15  & 0.20  & 0.15  &\nodata&\nodata&\nodata& 0.15  &\nodata&\nodata&\nodata&\nodata& 0.15 \\
HD 140283           & 0.10  & 0.10  & 0.20  & 0.10  & 0.25  & 0.10  &\nodata&\nodata& 0.10  &\nodata& 0.10  &\nodata&\nodata&\nodata&\nodata& 0.05 \\
HD 175305           &\nodata&\nodata&\nodata&\nodata&\nodata&\nodata&\nodata&\nodata&\nodata&\nodata& 0.15  &\nodata&\nodata& 0.25  & 0.15  & 0.20 \\
HD 196944           & 0.10  & 0.10  &\nodata& 0.10  & 0.05  & 0.10  & 0.15  &\nodata&\nodata&\nodata&\nodata&\nodata&\nodata&\nodata&\nodata& 0.15 \\
HD 222925           & 0.15  & 0.15  & 0.15  & 0.10  & 0.10  & 0.10  & 0.15  &\nodata& 0.15  &\nodata&\nodata& 0.20  & 0.20  &\nodata& 0.20  & 0.10 \\
\enddata
%\tablecomments{%
%}
\end{deluxetable*}

\begin{deluxetable*}{lcccccccccc}
\tablecaption{Mean Abundances
\label{abundtab}}
%\tablewidth{0pt}
\tabletypesize{\small}
\tablehead{
\colhead{Star} &
\multicolumn{2}{c}{Al~\textsc{i} (resonance)} &
\colhead{} &
\multicolumn{2}{c}{Al~\textsc{i} (high exc.)} &
\colhead{} &
\multicolumn{2}{c}{Al~\textsc{ii}} &
\colhead{[Fe/H]\tablenotemark{a}} &
\colhead{[Al/Fe]\tablenotemark{b}} \\
\cline{2-3} \cline{5-6} \cline{8-9}
\colhead{} &
\colhead{$\log\varepsilon$} &
\colhead{N} &
\colhead{} &
\colhead{$\log\varepsilon$} &
\colhead{N} &
\colhead{} &
\colhead{$\log\varepsilon$} &
\colhead{N} &
\colhead{} &
\colhead{} 
}
\startdata
BD $+$03\degree740  & 2.93 $\pm$ 0.10 & 4     & & \nodata         &\nodata& & 3.52 $\pm$ 0.17 & 1 & $-$2.93 $\pm$ 0.08 & $+$0.00 $\pm$ 0.17 \\
BD $+$44\degree493  & 2.09 $\pm$ 0.13 & 3     & & \nodata         &\nodata& & 2.92 $\pm$ 0.22 & 1 & $-$3.86 $\pm$ 0.16 & $+$0.33 $\pm$ 0.13 \\
%BD $-$13\degree3442 & 3.01 $\pm$ 0.10 & 3     & & \nodata         &\nodata& & 3.36 $\pm$ 0.20 & 1 & $-$2.85 $\pm$ 0.10 & $-$0.24 $\pm$ 0.20 \\
%CD $-$33\degree1173 & 2.90 $\pm$ 0.09 & 2     & & \nodata         &\nodata& & 3.30 $\pm$ 0.25 & 1 & $-$3.10 $\pm$ 0.09 & $-$0.05 $\pm$ 0.26 \\
%HD 2454             & 5.38 $\pm$ 0.18 & 1     & & 5.76 $\pm$ 0.15 & 2     & & 5.77 $\pm$ 0.20 & 1 & $-$0.35 $\pm$ 0.14 & $-$0.33 $\pm$ 0.20 \\
HD 19445            & 3.74 $\pm$ 0.10 & 4     & & \nodata         &\nodata& & 4.25 $\pm$ 0.10 & 1 & $-$2.18 $\pm$ 0.07 & $-$0.02 $\pm$ 0.08 \\
HD 84937            & 3.56 $\pm$ 0.10 & 9     & & \nodata         &\nodata& & 4.20 $\pm$ 0.10 & 1 & $-$2.28 $\pm$ 0.10 & $+$0.03 $\pm$ 0.11 \\
HD 94028            & 4.72 $\pm$ 0.11 & 6     & & \nodata         &\nodata& & 5.18 $\pm$ 0.20 & 1 & $-$1.56 $\pm$ 0.09 & $+$0.29 $\pm$ 0.20 \\
%HD 107113           & 5.37 $\pm$ 0.18 & 1     & & 5.70 $\pm$ 0.17 & 2     & & 5.70 $\pm$ 0.20 & 1 & $-$0.43 $\pm$ 0.15 & $-$0.32 $\pm$ 0.20 \\
HD 108317           & 3.25 $\pm$ 0.14 & 7     & & \nodata         &\nodata& & 3.90 $\pm$ 0.15 & 1 & $-$2.35 $\pm$ 0.12 & $-$0.20 $\pm$ 0.15 \\
HD 128279           & 3.26 $\pm$ 0.13 & 6     & & \nodata         &\nodata& & 3.72 $\pm$ 0.15 & 1 & $-$2.45 $\pm$ 0.10 & $-$0.28 $\pm$ 0.15 \\
HD 140283           & 2.94 $\pm$ 0.12 & 8     & & \nodata         &\nodata& & 3.65 $\pm$ 0.11 & 1 & $-$2.59 $\pm$ 0.08 & $-$0.21 $\pm$ 0.09 \\
HD 175305           & 4.21 $\pm$ 0.17 & 1     & & 4.97 $\pm$ 0.17 & 2     & & 4.76 $\pm$ 0.20 & 1 & $-$1.36 $\pm$ 0.10 & $-$0.33 $\pm$ 0.20 \\
HD 196944           & 3.22 $\pm$ 0.16 & 6     & & \nodata         &\nodata& & 4.16 $\pm$ 0.18 & 1 & $-$2.41 $\pm$ 0.08 & $+$0.12 $\pm$ 0.19 \\
HD 222925           & 4.23 $\pm$ 0.11 & 8     & & 4.79 $\pm$ 0.22 & 3     & & 4.77 $\pm$ 0.13 & 1 & $-$1.46 $\pm$ 0.11 & $-$0.22 $\pm$ 0.13 \\
\enddata
\tablenotetext{a}{Derived from Fe~\textsc{ii} lines}
\tablenotetext{b}{Derived from Al~\textsc{ii} and Fe~\textsc{ii} lines}
%\tablecomments{%}
\end{deluxetable*}

\section{Results}
\label{results}

Table~\ref{abundtab} lists the 
weighted mean \logeps{Al} abundances and [Al/Fe] ratios,
based on Al~\textsc{ii} and Fe~\textsc{ii} lines.
For these 11~stars spanning
$-3.9 <$~[Fe/H]~$< -1.3$,
the weighted mean [Al/Fe] is $-$0.06 $\pm$~0.04 ($\sigma =$ 0.22).
If two well-known carbon-enhanced stars
(\object[BD+44 493]{BD~$+$44$^{\circ}$493} and
\object[HD 196944]{HD~196944}; see, e.g.,
\citealt{ito13} and \citealt{placco15cemps})
are excluded,
the weighted mean [Al/Fe] ratio spanning
$-3.0 <$~[Fe/H]~$< -1.3$ is
$-$0.10 $\pm$~0.04 ($\sigma =$ 0.18).

\subsection{Comparison of Al Abundances from Different Lines}
\label{comparelines}

\begin{figure*}
\begin{center}
\includegraphics[angle=0,width=5.5in]{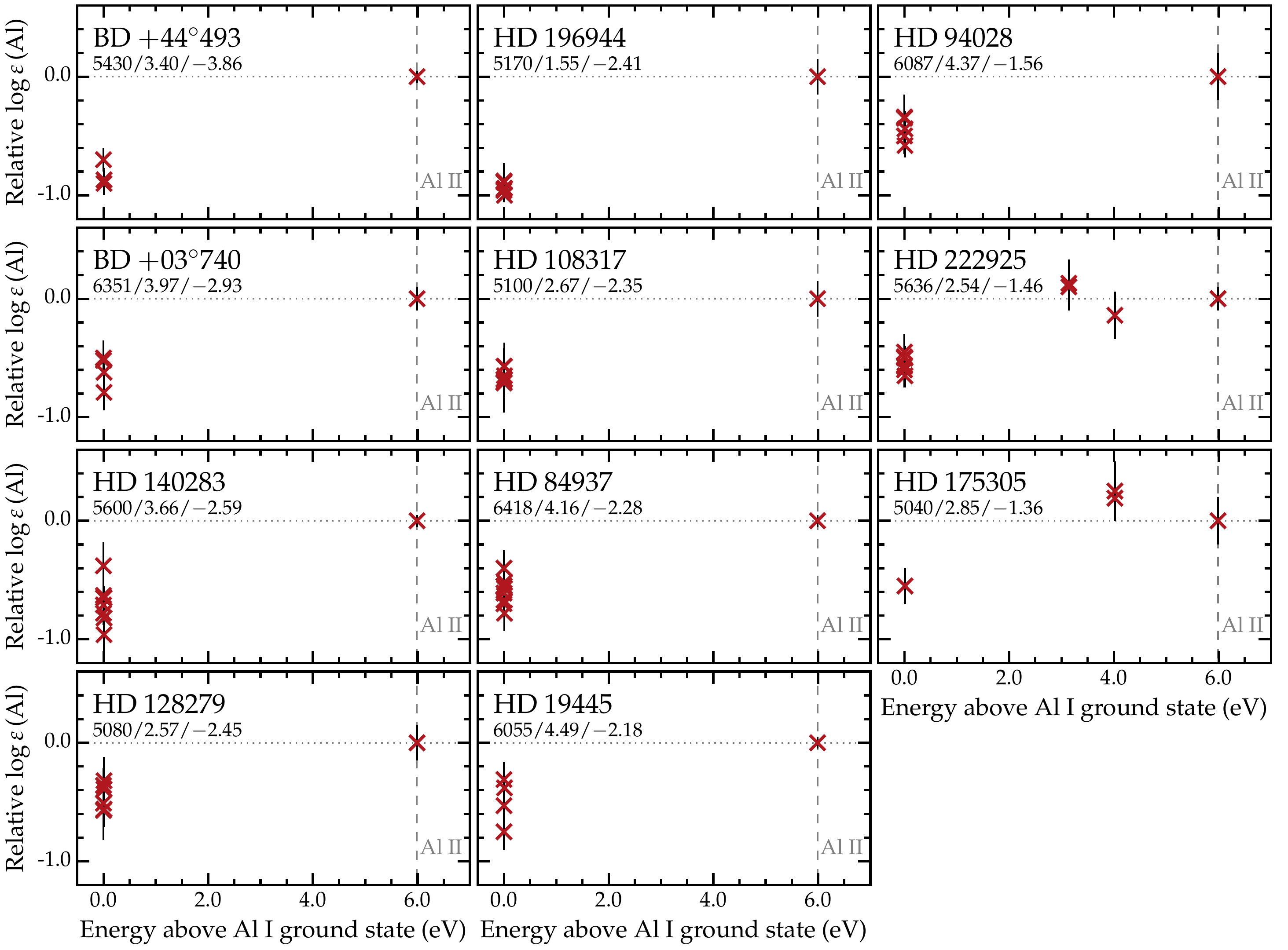}
\end{center}
\caption{
\label{deltaplot1}
Comparison of Al abundances derived from different lines.
The crosses mark abundances derived from individual lines.
Each set of abundances is scaled relative to the abundance
derived from the Al~\textsc{ii} line at 2669~\AA,
whose value is indicated as a dotted line at zero.
The first ionization potential of Al, 5.99~eV, is 
indicated by the vertical dashed line.
The \teff/\logg/[Fe/H] values
(in units of K, $\log$~cm~s$^{-2}$, and dex)
are listed.
The panels are ordered by increasing [Fe/H].
}
\end{figure*}

Figure~\ref{deltaplot1} illustrates the relative Al abundances
derived from different lines.
The Al~\textsc{i} resonance lines
yield consistent abundances in each star.
The high-excitation Al~\textsc{i} lines 
and the Al~\textsc{ii} line
also yield consistent abundances in each star.
These lines yield different abundances, however,
in the sense that the abundances derived from
Al~\textsc{i} resonance lines are lower than those
derived from high-excitation Al~\textsc{i} lines and the Al~\textsc{ii} line.

\begin{figure}
\begin{center}
\includegraphics[angle=0,width=3.35in]{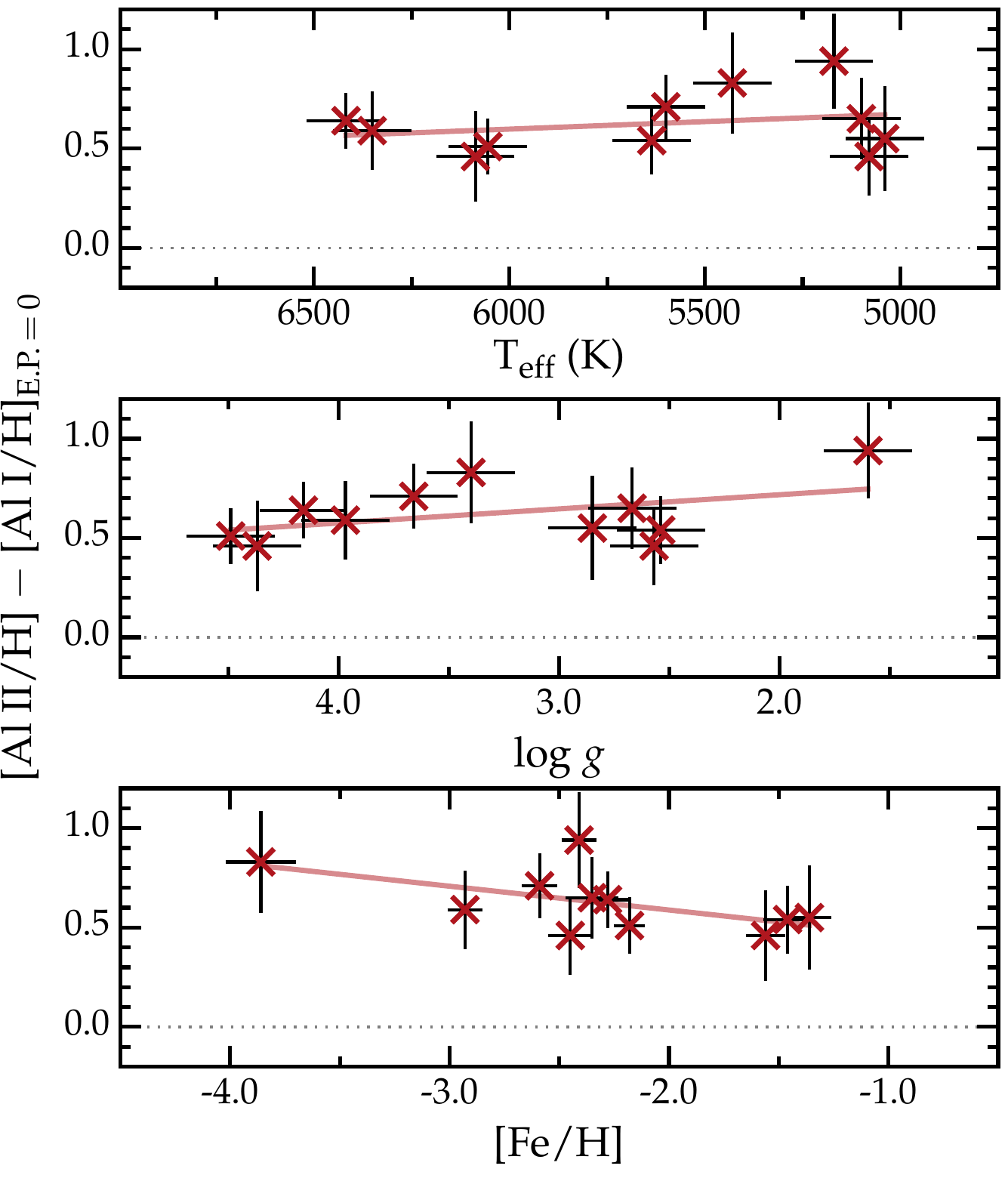}
\end{center}
\caption{
\label{deltaplot2}
Comparison of Al abundance differences 
between those derived from the Al~\textsc{ii} line at 2669~\AA\
and the Al~\textsc{i} resonance lines.
Linear fits are shown.
The dotted line marks a difference of zero.
}
\end{figure}

Figure~\ref{deltaplot2} illustrates these differences,
defined as the [Al/H] ratio derived from the Al~\textsc{ii} line
minus the [Al/H] ratio derived from the Al~\textsc{i} resonance lines,
as a function of \teff, \logg, and [Fe/H].
The differences range from $\approx +0.4$ to $+0.9$~dex.
The minimal differences between the coolest and warmest stars
in our sample
recalls earlier results by \citet{ryan96}, 
\citet{norris01}, and \citet{andrievsky08},
where the [Al/Fe] trends exhibited similar behavior in both dwarfs and giants.

\subsection{Comparison of LTE Al Abundances with Previous Studies}
\label{compareprevious}

Our stellar sample overlaps with two modern NLTE Al abundance studies.
We compare their LTE abundances, derived from the Al~\textsc{i} 
resonance lines, to ours, as a consistency check of our results.
We also compare our LTE Al abundance with that derived 
previously for the most metal-poor star in our sample.

There are two stars in common between our study and 
\citet{zhao16},
\object[HD 84937]{HD~84937} and 
\object[HD 94028]{HD~94028}.
Our LTE [Al/H] ratios differ from theirs by $+$0.04 and $+$0.21~dex,
respectively.
We have no lines in common with \citeauthor{zhao16} %.
If we rederive Al abundances from the UV Al~\textsc{i} resonance lines
using the \citeauthor{zhao16}\ model atmosphere parameters,
our [Al/H] ratios agree to within 0.02~dex and 0.10~dex.
Both values are within the stated abundance uncertainties.

There are also two stars in common between our study and
\citet{nordlander17al},
\object[HD 84937]{HD~84937} and
\object[HD 140283]{HD~140283}.
Our 1D LTE Al abundances derived from the resonance lines
differ from theirs by $+$0.09 and $-$0.20~dex, respectively.
Our model atmosphere parameters are very similar for 
\object[HD 84937]{HD~84937},
and our Al abundances decrease by only $-$0.01~dex if we adopt
their model atmosphere parameters.
The resulting [Al/H] ratios differ by only $+$0.08~dex, which is
within the stated uncertainties.
For \object[HD 140283]{HD~140283},
we adopt a model atmosphere with a cooler \teff\ (5600~K) than
\citeauthor{nordlander17al}\ did (5777~K).
The warmer value is supported by interferometric measurements of the 
radius by \citet{karovicova18,karovicova20}, 
but we adopt the cooler value for consistency with our previous work.
The [Al/H] abundance difference shrinks to only $+$0.04~dex if we instead
adopt their model, and this difference is well within the uncertainties.

\object[BD+44 493]{BD~$+$44$^{\circ}$493} is the most metal-poor star
in our sample, and no NLTE Al abundances have been published
for this star.
\citet{ito09} derived an LTE abundance \logeps{Al} = 2.06
from the Al~\textsc{i} line at 3961~\AA,
which is in excellent agreement with the LTE abundance 
we derive from this line, \logeps{Al} = 2.05.
We confirm this abundance using two additional Al~\textsc{i} lines in the UV.~

In summary, our LTE Al abundances derived from Al~\textsc{i} lines
are in agreement with previous studies of the same stars.

\section{Discussion}
\label{discussion}

\subsection{A Test of NLTE Calculations}
\label{comparenlte}

Many previous studies, including
\citet{baumueller97}, \citet{andrievsky08},
\citet{menzhevitski12}, \citet{mashonkina16al}, and \citet{nordlander17al},
have discussed the impact of NLTE on neutral Al in metal-poor stars.
NLTE photoionization from the
ground level of neutral Al is responsible for the 
low LTE Al abundances derived from Al~\textsc{i} resonance lines.
The Al neutral atom high-excitation levels are more closely coupled
to the ground level of the ion by resonant charge transfer reactions
involving H$^{-}$, 
so lines from these levels are formed near LTE.~
There is some variation among NLTE calculations,
even those that use identical atomic and collisional data
\citep{belyaev13},
concerning the magnitude of the corrections 
(see, e.g., discussion in \citeauthor{nordlander17al}).
In general, however, for metal-poor stars
the NLTE abundance corrections to the resonance lines are
positive and large, $\approx +0.3$ to $+$0.8~dex
for \teff\ $>$~5000~K and/or \logg\ $<$~4.0,
depending on the exact combination of stellar parameters.
The NLTE abundance corrections to the high-excitation lines are 
usually small, $\approx 0.1$~dex or less.
Figures~14 and 15 of \citeauthor{nordlander17al}\ illustrate the
NLTE corrections from that study and 
several others for comparison.

Previous studies 
(e.g., \citealt{mashonkina16al})
have also shown that
the ground state of Al$^{+}$, which gives rise to the
Al~\textsc{ii} line studied here, is formed in LTE.~
We conclude that the difference between
the LTE Al abundance derived from the Al~\textsc{ii} line and 
the LTE Al abundance derived from the Al~\textsc{i} resonance lines,
shown in Figure~\ref{deltaplot2},
presents an independent estimate of the NLTE corrections.
Our results therefore indicate that the
NLTE calculations are approximately correct
for the stellar parameter range reflected in our sample
(5000 $<$ \teff\ $<$ 6500~K; 
1.5 $<$ \logg\ $<$ 4.5;
$-3.9 <$ [Fe/H] $< -1.3$).
For example, \citet{nordlander17al} calculated NLTE corrections to the 
Al~\textsc{i} line at 3961~\AA\ of
$\approx+0.5$ to $+$0.8~dex for metal-poor giants
(\teff\ = 5250~K; $-4.0 \leq$ [Fe/H] $\leq -1.3$)
or
$\approx+0.4$ to $+$0.6~dex for metal-poor turnoff stars
(\teff\ = 6250~K; $-4.0 \leq$ [Fe/H] $\leq -1.3$),
which match our results.
This agreement offers a gratifying confirmation of the NLTE predictions.

\subsection{The [Al/Fe] Ratio in Metal-Poor Stars}
\label{alfe}

\begin{figure}
\begin{center}
\includegraphics[angle=0,width=3.35in]{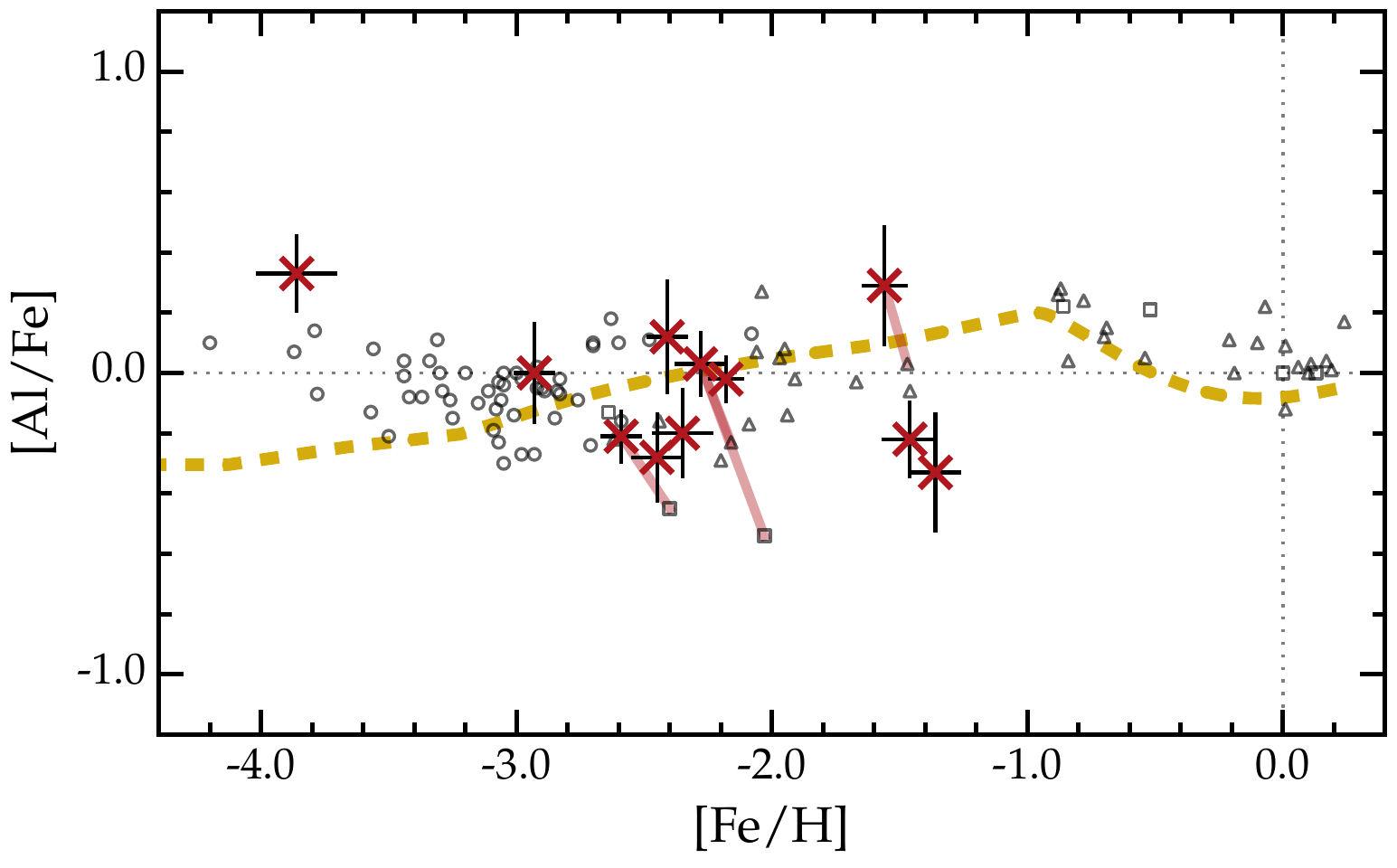}
\end{center}
\caption{
\label{alfeplot}
The [Al/Fe] ratio as a function of [Fe/H]
for the stars in our sample and several previous studies.
The large red crosses mark our 
abundances, derived from Al~\textsc{ii} and Fe~\textsc{ii} lines.
The 1D NLTE results from previous studies are indicated by
small gray circles \citep{andrievsky08},
small gray triangles \citep{zhao16}, and
small gray squares \citep{nordlander17al}.
The pink lines connect stars in common between our sample
and previous work.
The dashed yellow line marks the Galactic chemical evolution model
presented by \citet{kobayashi20}.
The dotted lines mark the solar ratios.
}
\end{figure}

Figure~\ref{alfeplot} illustrates the [Al/Fe] ratio
as a function of [Fe/H]
for the 11~stars in our sample.
NLTE [Al/Fe] ratios
derived from Al~\textsc{i} lines
\citep{andrievsky08,zhao16,nordlander17al}
are shown for comparison.
Figure~\ref{alfeplot} demonstrates that
our LTE [Al/Fe] ratios, derived from Al~\textsc{ii} and Fe~\textsc{ii} lines,
are in good agreement with the NLTE [Al/Fe] ratios
derived previously.

The [Al/Fe] ratio is roughly constant
and slightly subsolar across the metallicity range examined.
This result is in reasonable agreement with
predictions of Galactic chemical evolution models.
The best model of \citet{kobayashi20}, for example,
predicts only a modest rise in [Al/Fe] from $-$0.3 to $+$0.2 from
[Fe/H] $= -4$ to $-1$,
as shown in Figure~\ref{alfeplot}.

The mild tension appears most pronounced at the
lowest metallicities, where inhomogeneous chemical enrichment
could be expected to bias the results in small samples.
\object[BD+44 493]{BD~$+$44$^{\circ}$493}
is the most metal-poor star in our sample.
It is a carbon-enhanced metal-poor star
with no enhancement of neutron-capture elements
(CEMP-no star; e.g., \citealt{ito09,placco14bdp44}).
The [Al/Fe] ratio we derive using the Al~\textsc{ii} line,
$+$0.33 $\pm$~0.13,
is higher by $+$0.9~dex
than the [Al/Fe] ratio derived previously by \citeauthor{ito13},
$-$0.57 $\pm$~0.15.
We attribute this discrepancy to NLTE overionization of neutral Al.
Enhanced [Al/Fe] ratios are found in $\sim$50\% of CEMP-no stars,
and frequently they are correlated with enhanced 
[Na/Fe], [Mg/Fe], and [Si/Fe] ratios
(e.g., \citealt{norris13cemp}).
\citet{ito13} find similar levels of enhancement,
ranging from $+$0.30 to $+$0.49,
among these three ratios
in \object[BD+44 493]{BD~$+$44$^{\circ}$493}.
The super-solar [Al/Fe] ratio 
also improves the zero-metallicity supernova model fits to the
abundance pattern in 
\object[BD+44 493]{BD~$+$44$^{\circ}$493}
\citep{tominaga14,roederer16d}.
The enhanced [Al/Fe] ratio in 
\object[BD+44 493]{BD~$+$44$^{\circ}$493}
may not be representative, however,
of the [Al/Fe] ratio in extremely metal-poor stars
without excesses of carbon and other light elements.
New data based on the Al~\textsc{ii} line in
such stars would be welcome.

\section{Summary}
\label{summary}

We present Al abundances using, for the first time, an Al~\textsc{ii} line,
detected at 2669~\AA\ in HST/STIS spectra
of 11 metal-poor stars that span 
5000 $<$ \teff\ $<$ 6500~K,
1.5 $<$ \logg\ $<$ 4.5, and
$-3.9 <$ [Fe/H] $< -1.3$.
This line is formed in LTE, and
the Al abundances derived from this line,
and Fe abundances derived from Fe~\textsc{ii} lines,
yield [Al/Fe] ratios that are slightly sub-solar,
$-$0.06 $\pm$~0.04 ($\sigma =$ 0.22).
This value is in good agreement with previous studies
that have made use of NLTE calculations to derive
Al abundances from Al~\textsc{i} lines in optical and
infrared spectra.

The detection of this line enables a new test
of NLTE calculations of Al~\textsc{i} resonance lines
that are not formed in LTE.~
The differences between the Al abundance derived 
from this Al~\textsc{ii} line and the Al abundances derived
from Al~\textsc{i} resonance lines, $\approx+0.4$ to $+$0.9~dex, 
match the predicted NLTE corrections to LTE
abundances derived from the Al~\textsc{i} resonance lines.
The agreement is highly encouraging.
Whenever spectra covering 
the UV Al~\textsc{ii} line are unavailable---as is the case for
the vast majority of stars at present---the NLTE corrections to 
LTE abundances should be considered reliable
when applied on a line-by-line and star-by-star basis.

In future decades,
the UV Al~\textsc{ii} line at 2669~\AA\ will be an important 
abundance indicator of Al nucleosynthesis in the 
first generations of stars.
It lies in a clean spectral window in metal-poor stars.
It is detectable in stars with metallicity
at least as low as [Fe/H]~$\simeq -4$,
and it should be detectable when
[Al/H] $> -5$ in cool giants.
UV observations at this wavelength are feasible with STIS
and any of the future space missions 
with a high-resolution UV spectrograph
that have been proposed to NASA,
including
the Habitable Exoplanet Observatory
(HabEx; \citealt{gaudi20}),
the Large UV/Optical/Infrared Surveyor
(LUVOIR; \citealt{luvoir19}), and
the Cosmic Evolution Through UV Surveys mission
(CETUS; \citealt{heap19}).

\acknowledgments

We thank 
Chiaki Kobayashi for sharing her chemical evolution model predictions,
Shimon Kolkowitz for helpful discussions,
Ian Thompson for taking two of the MIKE spectra,
and the referee for a thoughtful report.
I.U.R.\ and J.E.L.\
acknowledge support provided by NASA through grants
GO-14765 and GO-15657
from STScI,
which is operated by the AURA
under NASA contract NAS5-26555.
I.U.R.\ also 
acknowledges financial support from
grant PHY~14-30152 (Physics Frontier Center/JINA-CEE)
and 
grant AST-1815403
awarded by the U.S.\ National Science Foundation (NSF).~
This research has made use of NASA's
Astrophysics Data System Bibliographic Services;
the arXiv pre-print server operated by Cornell University;
the SIMBAD and VizieR
databases hosted by the
Strasbourg Astronomical Data Center;
the ASD hosted by NIST;
the VALD, operated at Uppsala University, 
the Institute of Astronomy RAS in Moscow, 
and the University of Vienna;
the MAST at STScI; 
and the
IRAF software package.

\facility{%
ESO:3.6~m (HARPS),
HST (STIS), 
Keck (HIRES),
Magellan (MIKE),
McDonald:Smith (Tull Coud\'{e}),
VLT (UVES)}

\software{%
IRAF \citep{tody93},
matplotlib \citep{hunter07},
MOOG \citep{sneden73},
numpy \citep{vanderwalt11},
R \citep{rsoftware},
scipy \citep{jones01}
}

\bibliographystyle{aasjournal}
\bibliography{iuroederer}

\end{document}